\documentclass{article}
\usepackage{amsmath}
\usepackage{amsfonts}
\usepackage{amssymb}
\usepackage{graphicx}
\newtheorem{theorem}{Theorem}

\newtheorem{remark}[theorem]{Remark}

\begin{document}

\title{Towards Unstructured Mesh Generation\\Using the Inverse Poisson Problem}
\author{Guy Bunin}
\maketitle

\begin{abstract}
A novel approach to unstructured quadrilateral mesh generation for planar
domains is presented. Away from irregular vertices, the resulting meshes have
the properties of nearly conformal grids. The technique is based on a
theoretical relation between the present problem, and the inverse Poisson (IP)
problem with point sources. An IP algorithm is described, which constructs a
point-source distribution, whose sources correspond to the irregular vertices
of the mesh. Both the background theory and the IP\ algorithm address the
global nature of the mesh generation problem. The IP algorithm is
incorporated in a complete mesh generation scheme, which also includes an
algorithm for\ creating the final mesh. Example results are presented and discussed.

\end{abstract}

\section{Introduction}

\begin{figure}
[ptb]
\begin{center}
\includegraphics[
height=1.2213in,
width=1.7891in
]
{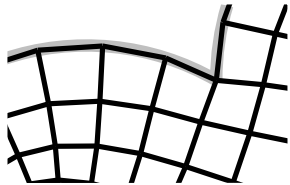}
\caption{Part of a boundary aligned mesh (black lines). The gray line
represents the domain's boundary.}
\label{fig:bound_align}
\end{center}
\end{figure}

\emph{Boundary alignment} is a critical feature of meshes in many
applications. In a boundary aligned mesh the boundary, or some other line, is
traced by the sides of high-quality cells, see fig. \ref{fig:bound_align}. The
definition of a \textquotedblleft well-shaped\textquotedblright\ cell may be
application dependent, but in many cases, cells similar in shape to squares
(for quadrilateral cells) or equilateral triangles (for triangles) are
preferred. Characteristics of the entire mesh are also important, such as
smooth cells-size and cell-shape transitions.

The problem of producing boundary aligned meshes with well-shaped cells has
been the subject of extensive research \cite{owen}. Still, many popular
algorithms are heuristic in nature, and a more general understanding of the
subject is called for, especially when quadrilateral meshes are considered.\ A
key difficulty is the problem's \emph{global} character: the shape and
position of every cell in the mesh is, at least in principle, related to that
of any other cell.

In a previous work \cite{CAGD}, we described a relation between the problem of
two-dimensional unstructured mesh generation, on both planar and curved
surfaces, and another well-known problem, namely the \emph{Inverse Poisson}
(IP) problem. The IP problem is concerned with reconstructing a \emph{source
distribution} $\rho$ of the Poisson equation $\nabla^{2}\phi=\rho$, from
information on the potential $\phi$ at the boundaries. In that work, the mesh
was assumed to be \emph{conformal} away from the irregular vertices (vertices
whose degree is different than four), like a grid mapped by a conformal
mapping. Such grids have the property of having square cells in the limit of
an increasingly finer grid. Under this assumption, the problem of mesh
generation was then shown to reduce to an IP problem. The irregular vertices
of the mesh correspond to point sources (delta functions) of $\rho$, and
$\phi$ is interpreted as the logarithm of the local resolution. This
theoretical framework turns the focus to the irregular vertices of the mesh:
once their distribution is fixed, the continuum properties of the mesh - local
resolution and directionality - are known. Note that it is also an explicitly
global formulation, since the resolution at any given point is affected, via
the function $\phi$, by the locations of all irregular vertices in the mesh.

In this paper the generation of planar quadrilateral meshes is discussed.
Resting on the results of \cite{CAGD}, a new IP algorithm is presented,
designed to construct source distributions of the appropriate type, which
approximate the resolution and mesh directionality inputs at user-specified
points, such as at the boundaries. The IP algorithm is then incorporated into
a complete mesh generation scheme, which also includes a technique for
generating the final mesh. An implementation is described, and shown in
example cases to generate boundary aligned meshes, where well-placed point
sources create smooth cell transitions and high quality cells. A similar
procedure is probably applicable to triangular meshes, but is not discussed in
the present work.

Remeshing of curved surfaces has recently attracted considerable attention;
for a review see \cite{alliez2005}. Many of the algorithms receive an input
mesh directionality throughout the surface, usually the principal curvature
directions of the surface. This setting presents different challenges than
those addressed here, since the mesh structure is determined, to a large
extent, when mesh directionality is given everywhere in the domain. For
example, the locations of critical irregular vertices are dictated by the mesh
directionality in the \emph{vicinity} of these points\footnote{In many works,
\emph{both} mesh directionality and local resolution are specified on the
entire surface, and since conformality requires a specific relation between
the two properties, the resulting meshes are not conformal (nor are they
claimed to be). In one exception \cite{ray}, a preprocessing step attempts to
adjust the local resolution, in order to create a more closely conformal
mesh.}. Another related subject is surface parameterization, concerned with
creating mapping of surfaces to the plane. Conformal surface parameterizations
are created in \cite{Gu}, but boundary alignment is not addressed.

The paper is organized as follows: Section \ref{sec:IP_review} shortly reviews
the relevant theoretical background, with emphasis on the relation between the
IP problem and unstructured mesh generation. Sections \ref{sec:algorithm} and
\ref{sec:create_final} describe the proposed IP algorithm, and the mesh
generation scheme. Section \ref{sec:implementation} describes an
implementation of the algorithm, and section \ref{sec:examples} gives examples
of meshes generated. Conclusions and possible directions for future research
are discussed in section \ref{sec:conclusions}.

\section{Background theory, relation to the IP problem\label{sec:IP_review}}

In this section an overview of the background theory will be given. Section
\ref{sec:motivation} explains the rationale underlying the mathematical
formulation. In section \ref{sec:definitions}, some key conformal geometry
concepts are discussed, together with their relevance to the present problem.
Section \ref{sec:IP_background} summarizes the results developed in
\cite{CAGD}, relating mesh generation with the IP problem. The exposition is
limited to planar two-dimensional mesh generation; A detailed account, in the
more general setting of curved surfaces, can be found in \cite{CAGD}.

\subsection{Motivation\label{sec:motivation}}

We start by considering mesh generation using conformal mappings, which can be
viewed as a special case of the theory to be described. Mapping techniques, in
general, construct a function from one domain, for which a mesh already
exists, to a second domain, which is being meshed. The mapping function is
then used to map the mesh into the second domain. A key idea is that continuum
properties of the mapping function control the shapes of the cells of the new
mesh, at least for small enough cells. For example, if the mapping is
conformal, i.e. angle preserving, a cell with right inner angles (rectangle)
will be mapped to a target cell with approximately right inner angles.

In unstructured mesh generation the connectivity of the mesh is not known in
advance, and a more general framework is called for. In what follows, the
interplay between the two domains which serves as a paradigm of mapping
techniques, is replaced with an interplay between two definitions of distances
on the input domain. Instead of imagining the mesh as being the image of a
mesh on a different domain, we \emph{redefine the distances} on the domain to
be meshed, and \emph{fix the cell edge length}. Thus, the local mesh
resolution will be proportional to the new local distance definition: a large
new distance between two given points will mean more cells will be placed in
between, hence a higher local resolution. Distances are redefined using the
concept of a \emph{metric}, known from Riemannian geometry.

Since we focus on cells which are squares in the limit of an increasingly
finer mesh, we need to define only two local properties: the resolution
(inverse of cell-size), and the direction of the\ cell-edges. The new
resolution, as noted before, is\ controlled by a new distance definition. We
use a new distance which is localy a scaling\ of the old distances: the new
distance between a point and other nearby points is equal to the old distance,
multiplied by a scalar factor which is independent of the direction. Thus, a
small square measured in one distance definition is also (approximately) a
square according to the new distance definition. In Riemannian geometry
terminology, the old and new metrics are said to be \emph{conformally
related}. The mesh directionality is related to the resolution, as is
described in the following section.

\subsection{Definitions, conformal geometry relations\label{sec:definitions}}

\begin{figure}
[ptb]
\begin{center}
\includegraphics[
height=2.0219in,
width=5.0384in
]
{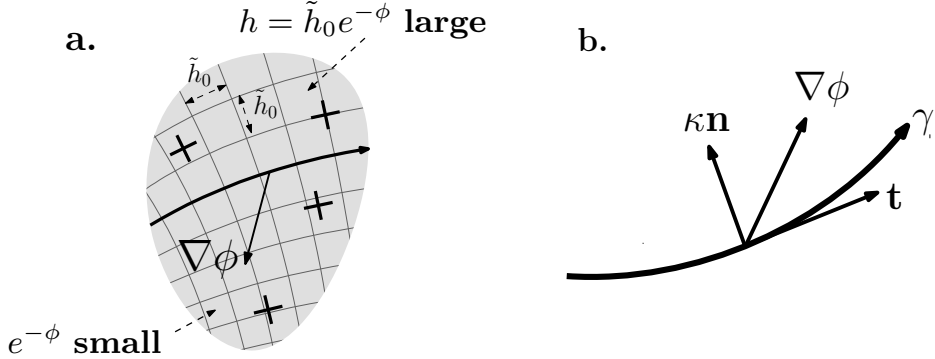}
\caption{\textbf{a.} A conformal grid. Solid lines are geodesics. The distance
between two parallel lines is $\widetilde{h_{0}}$, measured with the new
distance definition. Geodesics bend towards smaller cells. Crosses represent
the cross-field at selected points.\textbf{ b.} The curvature of a geodesic is
related to the gradient of $\phi$.}
\label{fig:lay_grid1}
\end{center}
\end{figure}

It is convenient to work with a function $\phi\left(  \vec{r}\right)  $,
defined as the \emph{logarithm} of the local scaling factor. That is, a small
square of side length $h$ measured with the original, Cartesian distance
definition, is a square with side length $\tilde{h}\left(  \vec{r}\right)
=h\left(  \vec{r}\right)  e^{\phi\left(  \vec{r}\right)  }$ as measured with
the new distance definition, see fig. \ref{fig:lay_grid1},a.. If we imagine
that the domain is covered with many small squares, all with the same side
length $\tilde{h}_{0}$ in the new distance definition, the size of these cells
in the original distance definition will be $h\left(  \vec{r}\right)
=\tilde{h}_{0}e^{-\phi\left(  \vec{r}\right)  }$. The (local) resolution is
the inverse of the local-cell-size, hence
\begin{equation}
resolution\propto e^{\phi\left(  \vec{r}\right)  },\label{eq:phi_is_log_res}
\end{equation}
where the proportionality constant is a single number for the whole mesh.

The second continuum property is the local directionality of the cell edges.
The edges should ideally meet at right angles everywhere, except at\emph{
irregular vertices}, where the number of edges incident on the vertex is
different than four. It is therefore natural to assign to every point a set of
four directions, mutually parallel or perpendicular. This concept was
expressed by many authors, and given various names, such as \emph{mesh
directionality} \cite{shimada}, \emph{4-symmetry direction field }\cite{nSym},
and \emph{frame field}, although the last may refer to a structure which\ also
holds cell-size information \cite{quadcover}. Graphically, this object can be
represented by a cross at every point, see fig. \ref{fig:lay_grid1},a., and
here will be called a \emph{cross-field }\cite{CAGD}. On the plane, the cross
direction can be measured by the angle $\theta$ from the $x$-axis to one of
the directions of the cross. This angle is fixed up to an addition of $\pi/2$
radians, i.e. $\theta_{1},\theta_{2}$ represent the same cross iff $\theta
_{1}=\theta_{2}+n\pi/2$, for some integer $n$.

The function $\phi$ and the cross-field are not unrelated; due to
conformality, lines that trace the edge directions bend towards the side with
smaller cells, see fig. \ref{fig:lay_grid1},a.. In the continuum theory, these
lines\ are known as \emph{geodesics} of the manifold. Geodesics are a
generalization of the concept of straight lines to non-Euclidian geometries.
In the original, Cartesian coordinates, the geodesic obeys the following
differential equation:
\begin{equation}
\kappa=\frac{\partial\phi}{\partial n},\label{eq:kappa_dphi_dn}
\end{equation}
where $\kappa$ is the curvature of the geodesic in the Cartesian coordinate
system, and $\partial\phi/\partial n$ is the derivative of the function $\phi$
in the direction normal to the tangent, see fig. \ref{fig:lay_grid1},b.. Eq.
(\ref{eq:kappa_dphi_dn}) allows one to calculate the change in the direction
of the cross-field between two points connected by a geodesic. There is also a
direct way of calculating the change in the direction of a cross-field between
any two points, along any curve connecting the two points, known as
\emph{parallel-transport}. Let $\alpha$ be some curve from point $a$ to point
$b$, and let $\theta_{a},\theta_{b}$ be the angles of the crosses at points
$a$,$b$ respectively (as noted above, the angles are defined up to an addition
of a $\pi/2$ radians). Then
\begin{equation}
\int_{\alpha}\frac{\partial\phi}{\partial n}ds=\theta_{b}-\theta_{a}
\text{.}\label{eq:theta_change_par_tranport}
\end{equation}
where the integration denotes a line integral along the curve $\alpha$,
according to the length parameter $s$ on $\alpha$, as measured in the original
coordinate system. The differential formulation of this equation is
\begin{equation}
\frac{\partial\phi}{\partial n}=\frac{\partial\theta}{\partial t}
,\label{eq:phi_theta_right_hand_relation}
\end{equation}
where $\left(  \vec{t},\vec{n}\right)  $ are a pair of perpendicular vectors
that form a right hand system.

Where the function $\phi$ is defined, that is, at any point in the domain that
is not a singularity, $\phi$\ can be shown to \emph{harmonic}, that is to obey
the Laplace equation:
\begin{equation}
\nabla^{2}\phi\equiv\frac{\partial^{2}\phi}{\partial x^{2}}+\frac{\partial
^{2}\phi}{\partial y^{2}}=0\text{,}\label{eq:phi_laplace}
\end{equation}
where, again, the derivatives are taken with the original coordinate system.
Eq. (\ref{eq:kappa_dphi_dn},\ref{eq:phi_laplace}) are well-known results of
conformal geometry \cite{Aubin},\cite{Chang}. Eq.
(\ref{eq:theta_change_par_tranport}) is derived in \cite{CAGD}

\subsection{Relation to the IP problem\label{sec:IP_background}}

Equations (\ref{eq:kappa_dphi_dn},\ref{eq:phi_laplace}) fully describe the
relations between the cross-field and the function $\phi$, at any regular
point of the domain, that is, any point that is not a singularity of the
function $\phi$. The singularities of the function $\phi$ are a key ingredient
of the theory, since they correspond to the irregular vertices of the mesh,
and unstructured meshes are those which contain irregular vertices. A detailed
analysis of the possible of singularities of the harmonic function $\phi$, and
their effect on the resulting mesh, was carried out in \cite{CAGD}, and shows
that the only type of singularity that corresponds to a mesh with a finite
number of cells is of the type $\phi\left(  \vec{r}\right)  \propto
\ln\left\vert \vec{r}-\vec{r}_{0}\right\vert $, where $\vec{r}_{0}$ is the
location of the singularity. In the IP literature, such a sigularity is known
as a \emph{point source}. Furthermore, the prefactor of the logarithm is
directly related to the degree of the irregular vertex in the final mesh. More
specifically, suppose there are $n_{c}$ singularities in a domain $D$, at
points $r_{m},m=1..n_{c}$, then the function $\phi$ can be written as

\textbf{Condition 1:}
\begin{equation}
\phi\left(  \vec{r}\right)  =\phi_{L}+
{\displaystyle\sum\limits_{m=1}^{n_{c}}}
\frac{Q_{m}}{2\pi}\ln\left\vert \vec{r}-\vec{r}_{m}\right\vert =\phi_{L}
+\frac{1}{4}
{\displaystyle\sum\limits_{m=1}^{n_{c}}}
k_{m}\ln\left\vert \vec{r}-\vec{r}_{m}\right\vert ,\label{eq:cond1}
\end{equation}
where $\phi_{L}$ is a harmonic function, $k_{m}\in
\mathbb{Z}
$, and $k_{m}>-4$. The numbers $Q_{m}$ are known as \emph{charges}, so
Condition 1 states that the charges are integer multiples of $\pi/2$. We will
refer to the numbers $k_{m}$ simply as the \textquotedblleft$k$
-values\textquotedblright\ of the singularities.\ The degree (number of
incident edges) of the irregular vertex corresponding to the source at
$\vec{r}_{i}$ is equal to $4+k_{m}$. So, for example, irregular vertices of
degrees 3 and 5 will correspond to a singularities with $k_{m}=-1$ and
$k_{m}=+1$, respectively.

In a geometrical context, a logarithmic singularity of $\phi$ represents a
\emph{cone-point} - the tip of a cone - in the new distance definition. The
charge corresponds to the angle deficit of the tip of the cone. Related
subjects where manifolds with cone-points are considered include the study of
disclinations in elastic media \cite{katanaev}, and surface parameterizations
\cite{kharevych}.

An unstructured mesh is required to be aligned with the boundary. Namely, that
one of the cross directions be parallel to the tangent to the boundary (where
a tangent exists):

\textbf{Boundary Alignment definition:}
\begin{equation}
\tan\left(  \theta\left(  p\right)  +n\frac{\pi}{2}\right)  =\frac{dy}
{dx}\left(  p\right)  ,\ \ \ p\in\Gamma,\label{eq:bound_align}
\end{equation}
where $\left(  y\left(  s\right)  ,x\left(  s\right)  \right)  $ is a curve
tracing the boundary, $p=\left(  x,y\right)  $ and $n\in
\mathbb{Z}
$. Since $\theta$ is fixed up to an addition of a multiple of $\pi/2$ radians,
this definition depends only on the cross-field itself, and not on the
particular choice of $\theta$.

\begin{figure}
[ptb]
\begin{center}
\includegraphics[
height=1.4726in,
width=3.2265in
]
{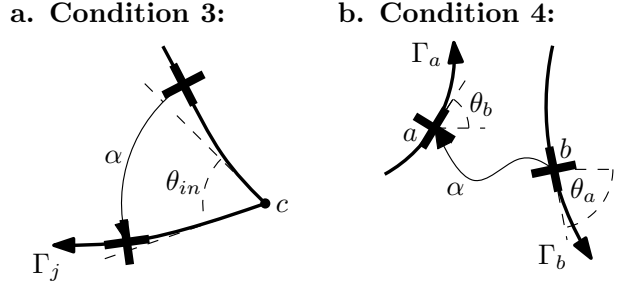}
\caption{Conditions 3 and 4.}
\label{fig:cond_3_4}
\end{center}
\end{figure}

Requiring boundary alignment, as defined in eq. (\ref{eq:bound_align}) can be
shown to be equivalent to three more conditions on the function $\phi$
\cite{CAGD} . The following, Condition 2, applies to a point on a smooth
section of the boundary, $\Gamma$. It roughly states that a smooth section of
the boundary is a geodesic. This provides information on the derivative of
$\phi$ normal to the boundary:

\textbf{Condition 2:}
\begin{equation}
\frac{\partial\phi}{\partial n}\left(  p\right)  =\kappa\left(  p\right)
,\ \ \ p\in\Gamma.\label{eq:cond2}
\end{equation}
$\kappa$ is the boundary curvature at $p$. The next condition is concerned
with \emph{junction points} of the boundary, where the tangent to the boundary
is discontinuous. It assures that the cross-field will be aligned with the
boundary on both sides of the junction point. Let $\alpha$ be a curve from
point $a$ to point $b$ on both sides of a point $c$ of the boundary, see fig.
\ref{fig:cond_3_4},b.. Then

\textbf{Condition 3:}
\begin{equation}
\int_{\alpha}\frac{\partial\phi}{\partial n}ds=n\frac{\pi}{2}+\theta_{in},
\label{eq:cond3}
\end{equation}
where $\theta_{in}$ is the angle of the boundary, and $n\in
\mathbb{Z}
$. This means that for some inner angles $\phi$ will contain a singularity at
$c$; at a distance $r$ from $c$, the singularity will be of type $\partial
\phi/\partial r\sim1/r$.

The final condition is concerned with the relation between two different
boundary components. It is just a restatement of eq.
(\ref{eq:theta_change_par_tranport}) for two points on two different boundary segments.

\textbf{Condition 4:}

Let $a$,$b$ be two points on two boundary curves $\Gamma_{a},\Gamma_{b}$. Let
$\alpha$ be a curve from $a$ to $b$. Then
\begin{equation}
\int_{\alpha}\frac{\partial\phi}{\partial n}ds=\theta_{b}-\theta_{a}
\text{.}\label{eq:cond4}
\end{equation}
Conditions 1-4 form a complete set of conditions on $\phi$, for a cross-field
that is boundary aligned to exist. That is, conditions 1-3, and condition 4 on
a number of selected curves, one from a selected point to each boundary
component, are sufficient for a boundary aligned cross-field to exist, see
\cite{CAGD}.

\section{Algorithm\label{sec:algorithm}}

\subsection{Mesh Generation Steps}

The mesh generation scheme consists of the following steps:

\begin{enumerate}
\item Setup and solution of the IP problem. Finds the sources' locations and
charges. Described in sections \ref{sec:input}-\ref{sec:calc_locations}.

\item Solution of the direct Poisson problem, to obtain the functions
$\phi,\theta$ throughout the domain, see section \ref{sec:restore_harmonic}.

\item Generation of the final mesh, see section \ref{sec:create_final}.
\end{enumerate}

\subsection{Input\label{sec:input}}

Suppose that we are given an open domain $D$, with boundary $\Gamma$, and are
given the required resolution function on $\Gamma$. Using eq.
(\ref{eq:phi_is_log_res}), this is readily translated to the required value of
$\phi$ on the boundary:
\begin{equation}
\left.  \phi\right\vert _{\Gamma}=\left.  \ln\left(  resolution\right)
\right\vert _{\Gamma}.\label{eq:phi_from_res}
\end{equation}

Condition 1 (eq. (\ref{eq:cond1})) is a solution of the Poisson equation
$\nabla^{2}\phi=\rho$, with point sources, whose locations and charges are yet
unknown. The problem is to find a distribution of sources (location and
charge) adhering to the boundary alignment definition, eq.
(\ref{eq:bound_align}), or alternatively to Conditions 2-4 (eq.
(\ref{eq:cond2})-(\ref{eq:cond4})), as well as to eq. (\ref{eq:phi_from_res}).
Such a problem is known as an\emph{ Inverse Poisson }(IP) problem. The IP
problem may be compared with the \emph{direct Poisson} problem, where the
source distribution is given, as well as some boundary information (e.g.,
Dirichlet or Newmann boundary conditions), and the value of the function
$\phi$ is to be found. In the IP problem, the source distribution is unknown,
and a source distribution adhering to the known boundary information is to be found.

IP problems have important applications in various areas of science and
engineering \cite{yamaguti},\cite{Hamalainen},\cite{Ioannides},\cite{johnston}
,\cite{zidarov}. By its nature, the IP problem is ill-posed, and the solution
might not be unique, and may be sensitive to small changes of the input, such
as small changes in boundary conditions. In delicate problems of this
type,\ any prior information on the source distribution may greatly affect the
applicability of a specific solution procedure. In the present problem, we
seek a point source distribution. A number of algorithms for solving an IP
problem with point sources appear in the literature. In \cite{ohe}, an inverse
problem where all sources have the same charge is solved. In our
implementation, however, at least two different charge values must be
incorporated ($Q=\pm\pi/2$). In \cite{el-badia},\cite{nara},\cite{baracharat}
an inverse problem where both the locations of sources and their charges are
unknown, and are reconstructed. This gives more freedom in the reconstruction
than we can allow, since for the present purposes the charges must be
multiples of $\pi/2$. Another important aspect of the present problem is that
the domain of the IP problem may be of any shape and topology (i.e., may
contain holes), whereas the above works only deal with a simply connected
region (a circle, usually).

It is important to note that in the present application, in contrast to other
standard applications, the input to the IP problem is not generated by some
existing source distribution (perhaps with some added noise), but by the
domain's shape and input resolution. The existence of a source distribution
which reconstructs the input data, at least approximately, is therefore not
obvious. This is an interesting and important subject, but is beyond the scope
of the present work.

\subsection{Complex Formulation\label{sec:complex_formulation}}

As is well known, the real and imaginary parts of a complex analytic function
are harmonic functions \cite{conway}. This correspondence has been utilized in
IP algorithms \cite{nara},\cite{el-badia},\cite{baracharat}, and will be used
here as well. We define the complex-valued function

\begin{equation}
F\left(  z\right)  =h\left(  z\right)  +\frac{1}{4}
{\displaystyle\sum\limits_{m=1}^{n_{c}}}
k_{m}\ln\left(  z-z_{m}\right)  , \label{eq:F_def}
\end{equation}
where $z_{m}=x_{m}+iy_{m}$, with $\left(  x_{m},y_{m}\right)  $ the components
of $\vec{r}_{m}$, that were defined in Condition 1. $h$ is a function on $D$,
such that
\[
\operatorname{Re}\left(  h\right)  =\phi_{L}.
\]
Then, recalling that $\operatorname{Re}\left(  \ln\left(  z\right)  \right)
=\ln\left\vert z\right\vert $, it follows that
\begin{equation}
\operatorname{Re}\left(  F\left(  z\right)  \right)  =\phi\left(  \vec
{r}\right)  , \label{eq:re_F_is_phi}
\end{equation}
with $\phi\left(  \vec{r}\right)  $ defined in Condition 1, and $z=x+iy$, with
$\vec{r}=\left(  x,y\right)  $.

The functions $h\left(  z\right)  $ and $\ln\left(  z-z_{m}\right)  $ for some
$z_{m}$\ are analytic in a neighborhood of any point that is not a singularity
of $\phi$. However, as functions over the entire domain $D$, they may be
\emph{multi-valued}.\ Multi-valued functions accept many-values at a point,
depending on the path taken to that point. It is well known that the complex
function $\ln\left(  z\right)  $ is multi-valued. Defining $\ln\left(
z\right)  $ as $\ln\left(  z\right)  \equiv\int_{\gamma}dt/t$, where $\gamma$
is a path from $1$ to $z$, the imaginary part of $\ln\left(  z\right)  $ is
only fixed up to an addition of multiples of $2\pi$, depending on the path
taken. If the domain $D$ is not simply-connected (i.e., contains holes), then
the function $h\left(  z\right)  $ may also be multi-valued, since by
following different paths around the holes, different values of $h\left(
z\right)  $ may be obtained.

The function $F\left(  z\right)  $, being a sum of multi-valued functions, may
also be multi-valued. According to eq. (\ref{eq:re_F_is_phi}), the real part
of $F$ is single-valued, as it is equal to $\phi$ at that point. We thus turn
to examine the imaginary part of $F$. The real and imaginary parts of a
complex function are related by the Cauchy-Reimann (CR) equations
\cite{conway}. Recall that for a complex differentiable function $f\left(
z\right)  =u\left(  z\right)  +iv\left(  z\right)  $, where $u\left(
z\right)  ,v\left(  z\right)  $ are real-valued, the Cauchy-Reimann equations
read $\partial_{x}u=\partial_{y}v$, and $\partial_{y}u=-\partial_{x}v$. This
allows one to recover the imaginary part of a analytic function if its real
part is given. Writing eq. (\ref{eq:phi_theta_right_hand_relation}) in the two
right hand coordinate systems $\left(  x,y\right)  ,\left(  y,-x\right)  $,
gives the relations $\partial\phi/\partial y=\partial\theta/\partial x$ and
$\partial\phi/\partial x=-\partial\theta/\partial y$. These are precisely the
CR equations for the complex function $\phi-i\left(  \theta+C\right)  $, where
$C$ is any real constant, hence
\begin{equation}
F\left(  z\right)  =\phi\left(  z\right)  -i\int_{\gamma}\frac{\partial\phi
}{\partial n}ds=\phi\left(  z\right)  -i\left(  \theta\left(  z\right)
+C\right)  \text{.}\label{eq:F_is_phi_i_theta}
\end{equation}
The constant $C$ is arbitrary, and will be taken to be zero. The multi-valued
nature of $F$ is explicit in the integral formulation in eq.
(\ref{eq:F_is_phi_i_theta}). The right-hand-side (RHS) of eq.
(\ref{eq:F_is_phi_i_theta}) shows that the since the imaginary part of $F$ is
$-\theta\left(  z\right)  $, and $\theta$ must be fixed up to additions of
$\pi/2$, then the integral
\begin{equation}
\int_{\gamma}\frac{\partial\phi}{\partial n}ds
\end{equation}
must also be fixed up to additions of $\pi/2$. Indeed, this was shown to hold
if Conditions 1-3 (eq. (\ref{eq:cond1})-(\ref{eq:cond3})) hold, see
\cite{CAGD}.

To summarize, the problem is restated as finding a multi-valued complex
function $F$, given by eq. (\ref{eq:F_def}), and whose value at points on the
boundary of $D$ is:
\begin{equation}
F\left(  z\right)  =\phi\left(  z\right)  -i\theta\left(  z\right)  .
\end{equation}

In the following sections, an algorithm for solving the problem as it was here
restated is described. The IP algorithm first removes the contribution of
$h\left(  z\right)  $ from the boundary conditions, paying special attention
to the junction points, and then constructs the source distribution.

\subsection{Handling junction points\label{sec:junc_points}}

The function $\phi$, and hence the function $F$, may be singular at junction
points. Special procedures for addressing this behavior are taken as part of
two different steps of the algorithm:

\begin{description}
\item[a.] Choosing the input resolution near junction points.

\item[b.] Special treatment of junction points when $h$ is removed from the
boudary conditions.
\end{description}

The two subjects are discussed in the following subsections,
\ref{sec:input_near_junc} and \ref{sec:remove_harmonic_junc}.

\subsubsection{Choosing the input resolution near junction
points\label{sec:input_near_junc}}

Except in the special cases when $\theta_{in}$ is a multiple of $\pi/2$, the
behavior of $\phi$ at a small neighborhood in $D$ of a junction point is
singular, with the singularity at the junction point, see Condition 3 (eq.
(\ref{eq:cond3})). If the input resolution in a small neighbourhood of the
junction point does not match this singular behavior, solution of the IP
problem will feature sources at any distance from the junction, no matter how
small, resulting in a distribution with an infinite number of sources. In
practice, this means that an IP algorithm will cluster many sources near the
junction point, in a futile attempt to reconstruct the boundary conditions
there. To avoid this problem, we adjust the \emph{input} $\phi$ at a
neighborhood of the junction point. This adjustment should vanish rapidly at a
distance larger than 1-2 cell-sizes, so as to have little effect on the
original required resolution.

Condition 3 states that the gradient of $\phi$ near a junction point must
diverge as $\partial\phi/\partial n\sim1/r$, where $r$ is the distance from
the junction point. Such a flux is formed by a logarithmic singularity at the
junction point (see also \cite{CAGD}, section 7.1). Consider a singular source
term of the form $\phi\left(  r\right)  =\frac{Q_{J}}{2\pi}\ln\left\vert
r\right\vert $, where $r$ is the distance from the junction point. Then the
flux through a circular arc $\alpha_{r}$, at distance $r$ from the junction
point is (see also fig. \ref{fig:cond_3_4},a.):
\[
\int_{\alpha_{r}}\frac{\partial\phi}{\partial n}ds=-\frac{Q_{J}}{2\pi}
\theta_{in},
\]
where $\theta_{in}$ is the junction inner angle. Then by Condition 3,
$\int_{\alpha_{r}}\frac{\partial\phi}{\partial n}ds=k\frac{\pi}{2}+\theta
_{in}$, hence
\begin{equation}
Q_{J}=2\pi\left(  n_{J}\frac{\pi/2}{\theta_{in}}-1\right)
.\label{eq:junc_charge}
\end{equation}
$n_{J}$ is integer, which must be positive, otherwise a mesh with an infinite
number of cells will result (\cite{CAGD}, section 7.1). In fact, $n_{J}$ has a
simple interpretation: it is the number of cells incident upon the junction
point in the resulting mesh\footnote{During the creation of the final mesh
connectivity (section \ref{sec:create_final}) nearby cone-points may be
joined, and cone-points may be shifted to the boundary, depending on the final
cell-size. In such a case, the number of cells incident on a junction point,
as well as on other boundary points, may change.}. A reasonable choice for
$n_{J}$ is therefore
\begin{equation}
n_{J}=round\left(  \frac{\theta_{in}}{\pi/2}\right)
,\label{eq:junc_incidence}
\end{equation}
for which the inner angles of the cells incident on the junction point are
closest to $\pi/2$.

In order to restrict the effect of this correction of $\phi$ to a small region
in $D$, a source term with the opposite charge $-Q_{J}$ can be placed
\emph{outside} $D$, at a distance of about 1-2 edge lengths. See for example
the second and third examples in section \ref{sec:examples}.

\subsubsection{Treatment of junctions when removing the harmonic
part\label{sec:remove_harmonic_junc}}

In section \ref{sec:remove_harmonic} a method will be described for
calculating and removing the contribution of $h\left(  z\right)  $ to the
boundary conditions. At junction points, however, the function $h\left(
z\right)  $ may be singular: as $D$ is an open set, boundary singularities are
part of $h\left(  z\right)  $, so care should be taken when preforming the
calculations described in section \ref{sec:remove_harmonic}, especially in a
numerical implementation of the technique.

To avoid these problems, we subtract the junction singularities from the
boundary value of $F\left(  z\right)  $ before proceeding with removing
$h\left(  z\right)  $. Let $z_{j}^{J}$ be the locations of the junction
points, and $Q_{J_{i}}$ the charges as given by eq.\ (\ref{eq:junc_charge}
),(\ref{eq:junc_incidence}). The complex function corresponding to
$\ln\left\vert z\right\vert $ is $\ln\left(  z\right)  $, so the contribution
of the junctions is
\begin{equation}
-
{\displaystyle\sum_{_{i}}}
\frac{Q_{J_{i}}}{2\pi}\ln\left(  z-z_{j}^{J}\right)  \text{.}
\end{equation}

If a singularity is located on an inner boundary component, the logarithms
must contain a branch cut somewhere in $D$. In order to avoid adding a cut, we
add another term to every hole, equal to minus the total charge\ in each hole.
Denote the inner boundaries by $\Gamma_{2},..,\Gamma_{n_{bnd}}$, and let
$z_{2}^{hole},..,z_{n_{bnd}}^{hole}$ be arbitrary points inside the respective
holes. The following terms are therefore added to $F\left(  z\right)  $:
\begin{equation}
F\left(  z\right)  \rightarrow F\left(  z\right)  -
{\displaystyle\sum_{J_{i}}}
\frac{Q_{J_{i}}}{2\pi}\ln\left(  z-z_{j}^{J}\right)  +\sum_{m=2}^{n_{bnd}}
{\displaystyle\sum_{J_{i}\in\Gamma_{m}}}
\frac{Q_{J_{i}}}{2\pi}\ln\left(  z-z_{m}^{hole}\right)
\end{equation}
In this way, the additional terms form an analytic function in $D$, with no
branch cut, and are a part of $h\left(  z\right)  $. Since the contribution of
$h\left(  z\right)  $ is removed from $F\left(  z\right)  $ as described in
the next section, the exact form of this additional term, i.e. the choice of
points $z_{m}^{hole}$, does not affect the results.

\subsection{Removing the harmonic part\label{sec:remove_harmonic}}

In this section a technique for calculating $h\left(  z\right)  $ is
presented. After $h\left(  z\right)  $ is calculated, its contribution to
$F\left(  z\right)  $ at the boundary can be subtracted. To simplify notation,
let $F_{p}\left(  z\right)  $ denote the sum of logarithms in $F\left(
z\right)  $:
\begin{equation}
F_{p}\left(  z\right)  \equiv F\left(  z\right)  -h\left(  z\right)  =\frac
{1}{4}
{\displaystyle\sum\limits_{m=1}^{n_{c}}}
k_{m}\ln\left(  z-z_{m}\right)  \text{.}\label{eq:Fp_def}
\end{equation}

In the context of IP algorithms, the idea that the contribution of the
harmonic part to the solution can be removed from the boundary information was
suggested in \cite{baracharat}. There, an IP problem in the unit disk $B$ is
discussed. The value of the analytic part of $F\left(  z\right)  $ on the
boundary of $B$ can then be calculated by taking the Fourier transform of
$F\left(  z\right)  $, and leaving only the positive frequency components, as
can be shown by considering the Lourent series of $F\left(  z\right)  $ in the
unit disk. By its construction, this technique applies to functions defined on
the unit disk. It can be extended to other domains, if a conformal mapping of
the domain to the unit disk (which exists according to the Riemann mapping
theorem) is calculated. For our present purposes, since the domain $D$\ in
every problem input is different, using this technique would require
constructing a mapping to the unit disk for each domain being meshed
separately. Furthermore, the domains in our problem may contain holes, which
further complicates the matter. We therefore use an alternative approach,
which is now described. It is based on the Cauchy integral theorem, and is
similar to the \textquotedblleft sum factorization\textquotedblright\ step in
the Weiner-Hopf technique \cite{noble}, applied to a bounded domain ($\Omega$,
defined below) in place of an infinite strip.

\begin{figure}
[ptb]
\begin{center}
\includegraphics[
height=1.5411in,
width=4.721in
]
{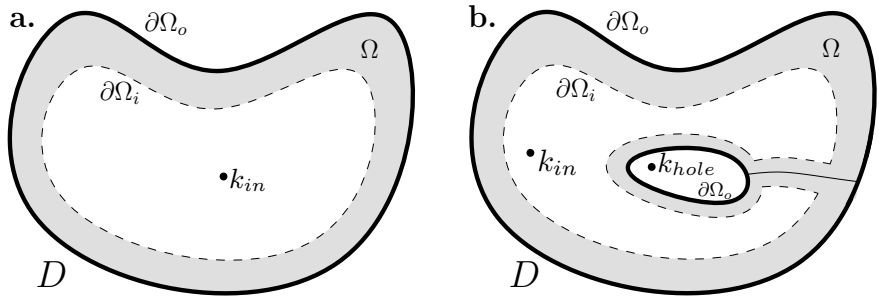}
\caption{Removing the harmonic part using the Cauchy integral. a. The case of
a simply-connected domain. b. A domain with more than one boundary component.
Thick line represents the boundary, thin line the cut introduced.}
\label{fig:cauchy}
\end{center}
\end{figure}

As in \cite{baracharat}, we assume there is a positive distance $d$\ between
the boundary of $D$ and the source closest to it, as is the case, for example,
when the number of sources is finite. Let $\Omega$ be the set of points at a
distance smaller than $d$ to the boundary, see fig. \ref{fig:cauchy},a.. As
will become apparent later, the value of $d$ does not enter the calculation
and is irrelevant, as long as there exists some $d>0$ as required.

We first consider a simply-connected domain $D$, with boundary $\Gamma$.
Define the flux through any curve $\alpha$:
\begin{equation}
\Phi_{\alpha}\equiv\int_{\alpha}\frac{\partial\phi}{\partial n}
ds.\label{eq:bound_flux_def}
\end{equation}
The flux $\Phi_{\Gamma}$ through the boundary can be calculated: the value of
$\partial\phi/\partial n$ at smooth boundary points is known from Condition 2
(eq. (\ref{eq:cond2})), and the flux at junction points is calculated from
Condition 3, as is explained in section \ref{sec:input_near_junc} above. This
flux can be shown to be a multiple of $\pi/2$.

\begin{remark}
\label{remark:flux_calc}Using Conditions 2,3 and the fact that the rotation of
a tangent in a simple curve is $2\pi\,$, the total flux through a boundary
component $\Gamma_{m}$ can be shown to be equal to $s\frac{\pi}{2}
(-4-\sum_{J_{i}\in\Gamma_{m}}\left(  n_{J_{i}}-2\right)  )$, where $J_{i}$ are
the junction points, $n_{J_{i}}$ for each junction is given by eq.
(\ref{eq:junc_incidence}), and $s=+1,-1$ for inner and outer boundary
components, respectively. See also \cite{CAGD}.
\end{remark}

To proceed with the decomposition calculation, we assume that $\Phi_{\Gamma
}=0$, as calculated from eq. (\ref{eq:bound_flux_def}) above. If this is not
the case, following \cite{baracharat}, we subtract a source term to the
boundary conditions, centered at some point $z_{in}$ inside the domain:
\begin{equation}
F\rightarrow F-\frac{k_{in}}{4}\ln\left(  z-z_{in}\right)
\label{eq:source_in_D}
\end{equation}
such that $\Phi_{\Gamma}$ is zero . As will be shown below, the choice of
$z_{in}$ does not affect the results. Since $\Phi_{\Gamma}$ was equal to a
multiple of $\pi/2$ before, $k_{in}\in
\mathbb{Z}
$. When $\Phi_{\Gamma}=0$, the function $F\left(  z\right)  $ is single-valued
and analytic in $\Omega$, as follows from the integral representation of
$F\left(  z\right)  $ in eq. (\ref{eq:F_is_phi_i_theta}), along with
$\Phi_{\Gamma}=0$. We now use the Cauchy's theorem, stating that for every
point $a\in\Omega$:
\begin{equation}
F\left(  a\right)  =\frac{1}{2\pi i}\int_{\partial\Omega}\frac{F\left(
z\right)  }{z-a}dz\text{.}\label{eq:cauchy_D_simply_con}
\end{equation}
The boundary $\partial\Omega$ of $\Omega$ has two boundary connectivity
components: the outer boundary (which is also the boundary of $D$), and the
inner boundary. Denote them by $\partial\Omega_{o}$ and $\partial\Omega_{i}$
respectively, see fig. \ref{fig:cauchy},a.. Eq. (\ref{eq:cauchy_D_simply_con})
now reads
\begin{equation}
F\left(  a\right)  =\frac{1}{2\pi i}\int_{\partial\Omega_{o}}\frac{F\left(
z\right)  }{z-a}dz+\frac{1}{2\pi i}\int_{\partial\Omega_{i}}\frac{F\left(
z\right)  }{z-a}dz\text{.}\label{eq:cauchy_on_omega}
\end{equation}
The function $\operatorname{Re}\left(  F\right)  =\phi$ is harmonic on
$\Omega$. According to a decomposition theorem (see \cite{HFT}, Chapter 9) its
decomposition into $\operatorname{Re}\left(  F_{p}\right)  $ and
$\operatorname{Re}\left(  h\right)  =\phi_{L}$, is also unique. It follows
that the decomposition of the multi-valued complex function $F$ into $F_{p}$
and $h$ is unique, up to the arbitrary constant $C$ in eq.
(\ref{eq:F_is_phi_i_theta}), chosen before to be zero. The two integrals
expressions on the RHS of eq. (\ref{eq:cauchy_on_omega}) correspond exactly to
the two components of the (unique) decomposition of $F\left(  z\right)  $ in
$\Omega$ as described in \cite{HFT}, hence
\begin{equation}
h\left(  z\right)  =\frac{1}{2\pi i}\int_{\Gamma}\frac{F\left(  z\right)
}{z-a}dz;\ \ F_{p}\left(  z\right)  =\frac{1}{2\pi i}\int_{\partial\Omega_{i}
}\frac{F\left(  z\right)  }{z-a}dz\text{,}\label{eq:cauchy_2_parts}
\end{equation}
where we have used the fact that $\Gamma=\partial\Omega_{o}$. We would like to
find $F_{p}$ on $\Gamma$, but $\partial\Omega_{i}$ is not known, nor is
$F\left(  z\right)  $ on $\partial\Omega_{i}$, so $F_{p}\left(  z\right)  $
cannot be computed directly from the second equation. The first equation can
however be used, since $F\left(  z\right)  $ is given on $\Gamma$, so
$h\left(  z\right)  $ can be calculated on $\Gamma$ (more precisely, since $h$
is defined in $D$, it is the limit of $h$ as $\Gamma$ is approached). Then
$F_{p}\left(  z\right)  $ on $\Gamma$ is given by $F_{p}\left(  z\right)
=F\left(  z\right)  -h\left(  z\right)  $, according to eq. (\ref{eq:Fp_def}).

Finally, we add back the source term subtracted before
\begin{equation}
F_{p}\rightarrow F_{p}+\frac{k_{in}}{4}\ln\left(  z-z_{in}\right)
.\label{eq:source_in_D_back}
\end{equation}

We now turn to the case when $D$ is not simply-connected. We add source terms:
one inside the domain, as discussed above (eq. (\ref{eq:source_in_D})), and
one inside every hole (i.e. \emph{outside} $D$), such that the flux
$\Phi_{\Gamma_{m}}$ through every boundary connectivity element is zero, see
Fig. \ref{fig:cauchy},b..
\begin{equation}
F_{p}\rightarrow F_{p}-\sum_{m=2}^{n_{bnd}}\frac{k_{hole_{i}}}{4}\ln\left(
z-z_{m}^{hole}\right)
\end{equation}
The $z_{m}^{hole}$ can be the same points used in section
\ref{sec:remove_harmonic_junc}, or other points. The results do not depend on
the additional charges' locations, as will be explained below. We now
introduce cuts so that the boundary contains a single connectivity element
(the cuts play a part in the derivation, but drop out of the final
calculation), see fig. \ref{fig:cauchy},b.. The cuts introduced also serve as
the branch cuts of the logarithms. Denote this new boundary $\widetilde
{\Gamma}$. Using $\widetilde{\Gamma}$, we proceed as when $D$ is simply
connected: define $\Omega$ as before, see fig. \ref{fig:cauchy},b., and
subtract the source term inside $D$, as in eq. (\ref{eq:source_in_D}).
According to eq. (\ref{eq:cauchy_2_parts}):
\begin{equation}
h\left(  z\right)  =\frac{1}{2\pi i}\int_{\widetilde{\Gamma}}\frac{F\left(
z\right)  }{z-a}dz=\frac{1}{2\pi i}\int_{\Gamma}\frac{F\left(  z\right)
}{z-a}dz+\frac{1}{2\pi i}\int_{cuts}\frac{F\left(  z\right)  }{z-a}
dz\text{.}\label{eq:cauchy_with_cuts}
\end{equation}
The second integral on the RHS denotes the integral over the cuts introduced
to form $\widetilde{\Gamma}$. Note that each cut-path is traversed twice, back
and forth. Since the flux $\Phi_{\Gamma_{m}}$ through each and every hole
boundary is zero, The value of $F\left(  z\right)  =\phi-i\int\partial_{n}\phi
ds$, when integrated along $\widetilde{\Gamma}$, is continuous across the
branch cuts, and the integrations over each cut traversed in both directions
cancel each other, and drop from the total integration. Therefore
\begin{equation}
h\left(  z\right)  =\frac{1}{2\pi i}\int_{\Gamma}\frac{F\left(  z\right)
}{z-a}dz\label{eq:cauchy_final_calc}
\end{equation}
as in eq. (\ref{eq:cauchy_2_parts}). As before, $F_{p}\left(  z\right)
=F\left(  z\right)  -h\left(  z\right)  $. Finally, the source term inside $D$
is added back, as in eq. (\ref{eq:source_in_D_back}). The boundary value of
$F_{p}$ obtained after this source term is added back may be multi-valued,
with a branch cut discontinuity in the imaginary part that is a multiple of
$\pi/2$. This is a valid input to the next step, as explained in section
\ref{sec:calc_locations} below (following eq. (\ref{eq:xsi_def})).

The choice of the added singularities' locations outside $D$ (inside the
holes) does not affect the results: changing the location of a singularity
with some charge\ from $z_{a}$ to $z_{b}$ is equivalent to adding a
singularity with opposite charge at $z_{a}$, and with the same charge at
$z_{b}$. Since these two singularities lay outside $\Omega$, and are of
opposite charge, this amounts to adding an analytic function to $F$, which is
removed by the Cauchy integral technique described above. The result is also
unaffected by the location $z_{in}$ of the source term subtracted inside $D$,
since its location only affects the $F_{p}$ and it is later added back. The
choice of cuts in eq. \ref{eq:cauchy_with_cuts}, of course, does not affect
the Cauchy integral calculation, since the cuts do not enter the final
calculation, eq. (\ref{eq:cauchy_final_calc}).

\subsection{Calculating the sources' locations\label{sec:calc_locations}}

At this point, we are given the value of $F_{p}\left(  z\right)  $ on the
boundary $\Gamma$ of the domain $D$. Using eq. (\ref{eq:F_def}), we define
\begin{equation}
\xi\left(  z\right)  \equiv\exp\left[  4F_{p}\left(  z\right)  \right]
=\exp\left[
{\displaystyle\sum\limits_{m=1}^{n_{c}}}
k_{m}\ln\left(  z-z_{m}\right)  \right]  \text{.}\label{eq:xsi_def}
\end{equation}
Note that while $F_{p}$ may be have a multi-valued imaginary part with a
branch cut discontinuity: $i\theta\rightarrow i\theta+in\pi/2$, for some
integer $n$, $\xi\left(  z\right)  $ is single valued, since $\exp\left[
4\left(  in\pi/2\right)  \right]  =1$.

Since the $k$-values, $k_{m}$, are integers, we can assume without loss of
generality that $k_{m}=\pm1$. Other charges may be formed by placing several
sources at the same location. We can therefore write eq. (\ref{eq:F_def}) as:
\begin{equation}
4F_{p}\left(  z\right)  =
{\displaystyle\sum\limits_{m=1}^{n_{+}}}
\ln\left(  z-z_{m}^{+}\right)  -
{\displaystyle\sum\limits_{m=1}^{n_{-}}}
\ln\left(  z-z_{m}^{-}\right)  ,\label{eq:fp_only_plus_minus}
\end{equation}
where $n_{+},n_{-}$ are the number of sources with $k_{m}=+1,-1$ respectively.

\begin{remark}
According to the divergence theorem, the difference $n_{+}-n_{-}$ is equal to
$\Phi_{\Gamma}/\left(  \pi/2\right)  $ (this can also be shown directly by
calculating the integral of $F_{p}$ along $\Gamma$). Since $\Phi_{\Gamma}$ is
known (see also remark \ref{remark:flux_calc}), $n_{+}-n_{-}$ is known.
Therefore, to fix $n_{+},n_{-}$ only one number remains to be chosen, e.g.
$n_{+}+n_{-}$. The choice of the total number of sources affects the accuracy
of the reconstruction of the input data, see sections \ref{sec:IPsummary} and
section \ref{sec:implementation}.\label{remark:divergence}
\end{remark}

Substituting eq. (\ref{eq:fp_only_plus_minus}), eq. (\ref{eq:xsi_def}) now
reads
\begin{equation}
\xi\left(  z\right)  =\frac{
{\displaystyle\prod\limits_{m=1}^{n_{+}}}
\left(  z-z_{m}^{+}\right)  }{
{\displaystyle\prod\limits_{m=1}^{n_{-}}}
\left(  z-z_{m}^{-}\right)  }\text{.} \label{eq:rational_roots}
\end{equation}

In eq. (\ref{eq:rational_roots}), the numbers $z_{m}^{+},z_{m}^{-}$ are the
roots of polynomials, and we proceed by presenting the polynomials in a
different form. Later, once the polynomials are found, the $z_{m}^{+}
,z_{m}^{-}$ are recovered by finding the roots of the polynomials. We rewrite
eq. (\ref{eq:rational_roots}) as
\begin{equation}
\xi\left(  z\right)  =\frac{
{\displaystyle\sum\limits_{m=0}^{n_{+}-1}}
p_{m}z^{m}+z^{n_{+}}}{
{\displaystyle\sum\limits_{m=0}^{n_{-}-1}}
q_{m}z^{m}+z^{n_{-}}},\label{eq:rational_standard_poly}
\end{equation}
where $p_{i},q_{i}$ are unknown coefficients. Note that the prefactors of
$z^{n_{+}},z^{n_{-}}$ are indeed equal to $1$, as can be seen by expanding the
polynomials in eq. (\ref{eq:rational_roots}). To find the unknown coefficients
numerically, we discretize eq. (\ref{eq:rational_standard_poly}), and give
$\xi\left(  z\right)  $ at $N$ points $z_{j}$, $j=1..N$, and write $\xi
_{j}\equiv\xi\left(  z_{j}\right)  $:
\begin{equation}
\xi_{j}=\frac{
{\displaystyle\sum\limits_{m=0}^{n_{+}-1}}
p_{m}z_{j}^{m}+z_{j}^{n_{+}}}{
{\displaystyle\sum\limits_{m=0}^{n_{-}-1}}
q_{m}z_{j}^{m}+z_{j}^{n_{-}}}.\label{eq:rational_discrete}
\end{equation}
The RHS of eq. (\ref{eq:rational_discrete}) is a \emph{rational function
interpolation} of the $\xi_{j}$ data (see e.g. \cite{recipies}). In the
implementation described below, the unknowns $p_{m},q_{m}$ are evaluated as
follows. Rearranging, eq. (\ref{eq:rational_discrete}) reads
\begin{equation}
\xi_{j}-z_{j}^{n_{+}-n_{-}}=
{\displaystyle\sum\limits_{m=0}^{n_{+}-1}}
p_{m}\left(  z_{j}^{m-n_{-}}\right)  -
{\displaystyle\sum\limits_{m=0}^{n_{-}-1}}
q_{m}\left(  \xi_{j}z_{j}^{m-n_{-}}\right)  .\label{eq:linear_system}
\end{equation}
These are $N$ linear equations for the $n_{+}+n_{-}$ unknowns: $p_{m},q_{m}$.
Since $N$ will typically be larger than $n_{+}+n_{-}$, the solution will only
be approximate, e.g., a solution in the least-mean-square (LMS) sense.

Once the unknown variables $p_{m},q_{m}$ are found, the source terms'
locations $z_{m}^{-},z_{m}^{+}$ are calculated by finding the roots of the two
polynomials appearing in the RHS of eq. (\ref{eq:rational_discrete}).

\begin{remark}
A solution in the LMS sense, as described in eq.\ (\ref{eq:linear_system}), is
not recommended if the resulting errors in the $\xi_{j}$'s are large, when
compared with $\xi_{j}$. This is due to two reasons: First, the input data
$\xi_{j}$ appears in both sides of eq. (\ref{eq:linear_system}), so the errors
in the LMS\ solution might not reflect the errors in $\xi_{j}$, and moreover,
the LMS\ solution may be sensitive to the choice of the coordinate system
origin. Secondly, an error in $\xi_{j}$ scales as a difference in the
resolution (since $\left\vert \xi_{j}\right\vert =\exp\left(  \phi\right)  $),
rather than the more \textquotedblleft natural\textquotedblright\ error
definition given by the \emph{ratio }of input and obtained resolution. This
means that at low resolutions (large cell-sizes), errors in $\xi$ might
represent large relative resolution deviations.
\end{remark}

\subsection{Restoring the harmonic part\label{sec:restore_harmonic}}

Once the sources' locations have been determined, the function $\phi$ can be
calculated everywhere in the domain. The solution will usually be approximate,
i.e., the resulting sum of sources will only approximate the required
$F_{p}=F-h$, and some deviation from both boundary alignment and resolution
requirements may be found. The trade-off between the two requirements can
be\ partly controlled by choosing an appropriate harmonic $\phi_{L}$ in eq.
(\ref{eq:cond1}). In order to satisfy the cell-size requirements exactly, one
can solve the (direct) Poisson equation, given by eq. (\ref{eq:cond1}), but
this time with a known charge distribution, and with Dirichlet boundary
conditions, i.e. by specifying $\phi$ on the boundary.

A different approach is to try and satisfy the boundary alignment conditions,
by solving eq. (\ref{eq:cond1}) with Newmann boundary conditions. This gives a
new $\phi$ on the boundary. The difference between the input (required) $\phi
$, and the $\phi$ obtained on the boundary can be used to evaluate the quality
of the source distribution calculated by the IP algorithm, and the total
number $n_{+}+n_{-}$ of sources can be changed accordingly. Note that when $D$
is not simply connected, solving $\phi_{L}$ with Neumann boundary conditions
(Condition 2) does \emph{not} mean that exact boundary alignment is obtained,
because Condition 4 might not be exactly fulfiled.

\subsection{Summary of the IP algorithm\label{sec:IPsummary}}

The steps followed in calculating the source distribution (locations and
charges) can be summarized as follows:

\begin{enumerate}
\item Adjusting\ the input resolution near the junction points (section
\ref{sec:input_near_junc}).

\item Subtracting the junction source terms (sections
\ref{sec:remove_harmonic_junc},\ref{sec:remove_harmonic}).
\begin{align*}
F  &  \rightarrow F-
{\displaystyle\sum_{\left\{  junctions\right\}  }}
\frac{Q_{J_{i}}}{2\pi}\ln\left(  z-z_{j}^{J}\right)  +\sum_{m=2}^{n_{bnd}}
{\displaystyle\sum_{J_{i}\in\Gamma_{m}}}
\frac{Q_{J_{i}}}{2\pi}\ln\left(  z-z_{m}^{hole}\right) \\
&  -\frac{k_{in}}{4}\ln\left(  z-z_{in}\right)  -\sum_{m=2}^{n_{bnd}}
\frac{k_{hole_{m}}}{4}\ln\left(  z-z_{m}^{hole}\right)  .
\end{align*}

\item Calculating the Cauchy integral to find $h\left(  z\right)  $ on the
boundary. Finding $F_{p}=F-h$ on the boundary. (section
\ref{sec:remove_harmonic}).

\item Adding back the source inside $D$\ (section \ref{sec:remove_harmonic}):
\[
F_{p}\rightarrow F_{p}+\frac{k_{in}}{4}\ln\left(  z-z_{in}\right)  .
\]

\item Calculating the locations of sources (section \ref{sec:calc_locations}),
using some initial total number of sources $n_{+}+n_{-}$.

\item Solving the (direct) Poisson problem to find the new $F$ approximating
the original $F$. According to the quality of reconstruction, step 5 can be
repeated with a different total number of charges.
\end{enumerate}

\section{Creating the final mesh\label{sec:create_final}}

\begin{figure}
[ptb]
\begin{center}
\includegraphics[
height=3.4895in,
width=4.5074in
]
{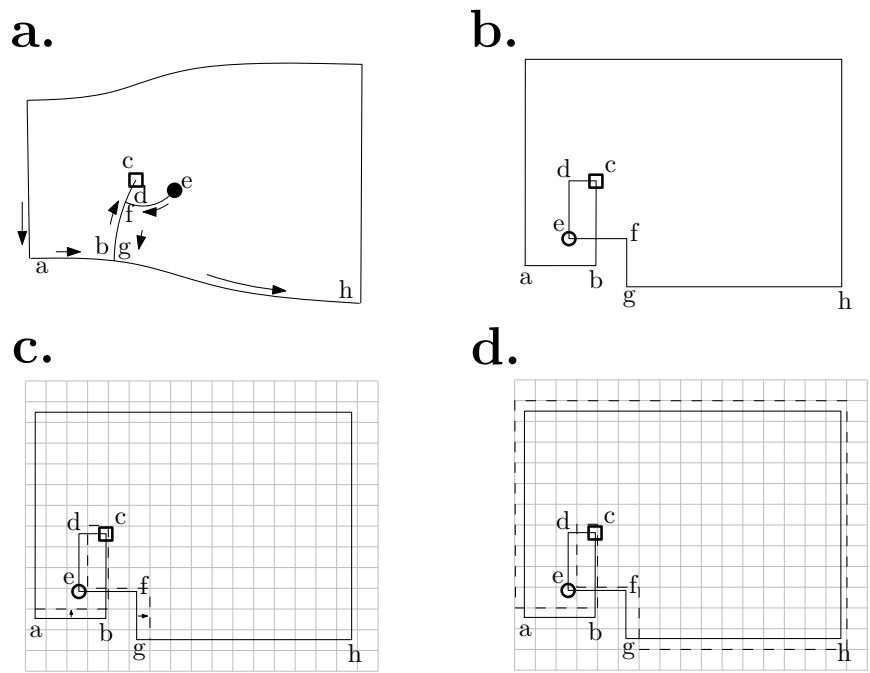}
\caption{Creating the final mesh. \textbf{a.} The cut-tree. Points $a,c$ are
source locations. \textbf{b.} The mapped cut-tree. \textbf{c.} Modifying the
boundary. Integer grid lines are drawn in gray. Sections $\overline
{ab},\overline{bc},\overline{cd},\overline{de}$ and $\overline{ef}$ are
shifted to the nearest grid line (dashed line). Section $\overline{fg}$ is
shifted so that $\overline{de}$ and $\overline{ef}$ will be of equal length.
\textbf{d.} Section $\overline{gh}$ is shifted so that the lengths
$\overline{bc}=\overline{cd}+\overline{fg}$. The process is continued until
point $a$ is reached again.}
\label{fig:mapping_sketch}
\end{center}
\end{figure}

Once the function $\phi\,$\ is set throughout the domain, the final mesh can
be constructed. There are various possible approaches to this problem. We
present a simple method that was used to create the examples in section
\ref{sec:examples}. In this method, the domain is cut along geodesics that
follow the cross-field directions. Every source is at the end of a cut, and
every inner boundary component is connected to the outer boundary by a cut,
see fig. \ref{fig:mapping_sketch},a.. The direction of the geodesics emanating
from the sources and directed along the cross-field are calculated as
explained in Appendix A. The cut domain, which we denote by$\ D^{\prime}$,
does not contain any sources in its interior, and its boundary consists of a
single connected component. A conformal mapping $g$ of $D^{\prime}$ into the
plane can be constructed, with a conformal factor (local scaling) of
$\exp\left(  \phi\right)  $.

We consider $D^{\prime}$ and $g\left(  D^{\prime}\right)  $ as lying in the
complex plane. First, note that since $D^{\prime}$ contains no sources in its
interior and is simply-connected, $\theta\left(  z\right)  $ is uniquely
defined in $D^{\prime}$. Denote by $\theta_{D^{\prime}}\left(  z\right)  $
this single-valued\ $\theta\left(  z\right)  $. To calculate the mapped
boundary $g\left(  D^{\prime}\right)  $, we calculate the function $g\left(
z\right)  $ along the (single) boundary $\Gamma^{\prime}$ of $D^{\prime}$:
\[
\left.  g\left(  z\right)  \right\vert _{\Gamma^{\prime}}=\int_{\Gamma
^{\prime}}\exp\left(  \phi\left(  z\right)  -i\theta_{D^{\prime}}\left(
z\right)  \right)  dz.
\]
\qquad

If this mapping is used to create a mesh, by placing a Cartesian grid on
$g\left(  D^{\prime}\right)  $, and mapping it back to $D^{\prime}$ by taking
the inverse of $g$, the mesh obtained will be invalid: the grid will\ not be
continuous across the cuts, will contain invalid cells at the sources, and cut
cells at the boundaries. To correct this, and obtain a valid final mesh, we
seek a function $f\left(  z\right)  $ on $\Gamma^{\prime}$ approximating
$g\left(  \Gamma^{\prime}\right)  $. A related problem, for surface meshes
without boundary alignment, was addressed in \cite{quadcover}.

First, note that because the cuts are added along geodesics aligned with the
cross field, $g\left(  \Gamma^{\prime}\right)  $ is a piecewise straight line,
whose straight segments are either horizontal or vertical. In order to obtain
a valid final mesh, the new function $f\left(  z\right)  $ should have the
following properties:

\begin{enumerate}
\item $f\left(  z\right)  $\ must map the closed path $\Gamma^{\prime}$ to a
closed path.

\item $f\left(  \Gamma^{\prime}\right)  $ must be a piecewise straight line,
composed of horizontal and vertical lines only. To each segment in $g\left(
\Gamma^{\prime}\right)  $ corresponds a segment in $f\left(  \Gamma^{\prime
}\right)  $, at the same order along the paths.

\item Each segment in $f\left(  \Gamma^{\prime}\right)  $ must be directed in
the same direction as the corresponding segment in $g\left(  \Gamma^{\prime
}\right)  $, or be of zero length.

\item If $\gamma^{\prime}$ is a geodesic cut of $\Gamma^{\prime}$, it is
followed in both directions: $\gamma^{\prime}$ and $\gamma^{\prime-}$. Then
$f\left(  \gamma^{\prime}\right)  $ and $f\left(  \gamma^{\prime-}\right)  $
must have the same length. This ensures the same number of cells on both sides
of a single cut.
\end{enumerate}

We describe a simple algorithm for creating $f\left(  \Gamma^{\prime}\right)
$, meeting these requirements. To create $f\left(  \Gamma^{\prime}\right)  $,
the path $g\left(  \Gamma^{\prime}\right)  $ is followed, and the segments of
$\left.  g\left(  z\right)  \right\vert _{\Gamma^{\prime}}$ are modified one
after the other, in the order they appear in $g\left(  \Gamma^{\prime}\right)
$, see fig. \ref{fig:mapping_sketch}. Each segment is shifted to a parallel
line of the integer grid. This is usually the nearest segment, except in the
case the length of the segment directly before this segment dictates a
different shift, according to properties (3,4) above, see fig.
\ref{fig:mapping_sketch},c.. This algorithm requires that at least one
junction exist in $\Gamma$, which is mapped to a right angle in $g\left(
\Gamma^{\prime}\right)  $. The algorithm starts from the segment following
this junction, so that when the last segment is reached and shifted, a change
in the length of the first segment will be allowed.

The resulting modified path is similar to $g\left(  \Gamma^{\prime}\right)
\,$, only with different segments lengths. $f\left(  \Gamma^{\prime}\right)  $
is then defined as the composition of $g\left(  \Gamma^{\prime}\right)  $
followed by a linear mapping of each segment of $g\left(  \Gamma^{\prime
}\right)  $ to the corresponding segment in the modified path. Once $f\left(
\Gamma^{\prime}\right)  $ is known, $f\left(  D^{\prime}\right)  $ can be
defined. In the implementation described below this is done by placing a
triangular mesh in $D^{\prime}$, and solving the Laplace equation for both
$\operatorname{Re}\left(  f\left(  z\right)  \right)  $ and $\operatorname{Im}
\left(  f\left(  z\right)  \right)  $ on $D^{\prime}$. $f\left(  z\right)  $
obtained in this way is not in general analytic, since the CR conditions might
not hold between $\operatorname{Re}\left(  f\left(  z\right)  \right)  $ and
$\operatorname{Im}\left(  f\left(  z\right)  \right)  $, so the final mesh
will not exactly conformal, but only approximately, as $f\left(  z\right)  $
approximates the analytical function $g\left(  z\right)  $. The mesh edges are
then extracted by tracing the integer-valued lines of $\operatorname{Re}
\left(  f\left(  z\right)  \right)  $ and $\operatorname{Im}\left(  f\left(
z\right)  \right)  $.

Once the connectivity of the mesh has been established, a final smoothing
procedure\ which involves all the interior vertices, including those on the
cut path, is preformed. In the definition of $f\left(  z\right)  $, both
$\operatorname{Re}\left(  f\left(  z\right)  \right)  $ and $\operatorname{Im}
\left(  f\left(  z\right)  \right)  $ are harmonic functions. A standard
finite difference approximation to a harmonic function on a uniform grid,
assigns to each vertex the average of its four neighboring vertices. This\ is
applied to all vertices inside the mesh, including those on the cuts of
$\Gamma^{\prime}$, which can be interpreted as a generalization of the
defintion of $f\left(  z\right)  $ (to be more precise, in this smoothing
procedure it is $f^{-1}\left(  z\right)  $ which is assumed to be harmonic).
This procedure is the well-known Laplacian smoothing procedure, in which the
location of every vertex is equal to the average location of its neighboring
vertices, and in this application it is theoretically justified if $f\left(
z\right)  $ is approximately $g\left(  z\right)  $, i.e. if $f\left(
\Gamma^{\prime}\right)  $ is close to $g\left(  \Gamma^{\prime}\right)  $.

\section{Implementation Details\label{sec:implementation}}

The algorithm described in the previous section was implemented in Matlab
\cite{matlab}, using standard Matlab functions, such as a least-mean square
solver for linear systems, an ODE solver, etc..

To calculate the Cauchy integral (section \ref{sec:remove_harmonic}), the
boundary was approximated by a piecewise linear path between sampled points,
with a constant value of $\phi$ on each segment. This allows the integral over
each segment of the path to be calculated analytically, and the path integral
there is equal to the sum of the integral over the segments. This is a low
order approximation, but was sufficient for the present purposes.

The linear system, eq. (\ref{eq:linear_system}), was solved in the
least-mean-square sense, using Matlab's \texttt{mldivide}\ function. Note that
the coefficients in this system contain powers of the $z_{j}$'s, the boundary
data locations. This leads to ill-conditioned linear systems, which are more
sensitive to the numerical round-off errors, as the number of sources
$n_{+},n_{-}$ increases, or the range of $\left\vert z\right\vert $ increases.
Tests indicate that when the solver issues a warning, indicating that the
matrix is rank deficient to the working precision, the results of the IP
algorithm become unreliable (this was not the case in the examples in section
\ref{sec:examples}). For input domains where many sources or a high
$\left\vert z\right\vert $ range are required, a different solver
implementation may be needed, such as a varying precision arithmetic
computation, working at higher floating-point accuracies.

After sources' locations are recovered by taking the roots of the polynomials,
the function $\phi_{L}$ of eq. (\ref{eq:cond1}) is calculated, given the
Newmann boundary conditions in eq. (\ref{eq:cond2}). This was done using the
Method of Fundamental Solutions (MFS), see e.g. \cite{MFS}, where the solution
to the Laplace for $\phi_{L}$ is approximated by a sum of (real) source terms,
i.e. functions of the form $\phi\left(  r\right)  =\frac{Q_{i}}{2\pi}
\ln\left\vert \vec{r}-\vec{r}_{i}\right\vert $, with the locations $\vec
{r}_{i}$ lying \emph{outside} the domain $D$\ where $\phi_{L}$ is harmonic.
The charges $Q_{i}$ are then calculated to best fit the boundary conditions.

Once the function $\phi$ is given everywhere in $D$, the directions of the
star-geodesics at each source are calculated, as described in Appendix A. The
star-geodesics, emanating from the source were then traced by solving the
differential equation, eq. (\ref{eq:kappa_dphi_dn}), using a Runge-Kutta ODE
solver. The cut-tree was constructed by taking one source at a time at some
arbitrary order, and introducing a cut along a star-geodesic of the added
source, that is closest to the cut-tree constructed so-far. This construction
of the cut tree is admitantly arbitrary, and may not be optimal in some cases.

Lastly, the algorithm described in section \ref{sec:create_final} is applied.

\section{Example Results\label{sec:examples}}

\begin{figure}
[ptbh]
\begin{center}
\includegraphics[
height=4.6769in,
width=6.249in
]
{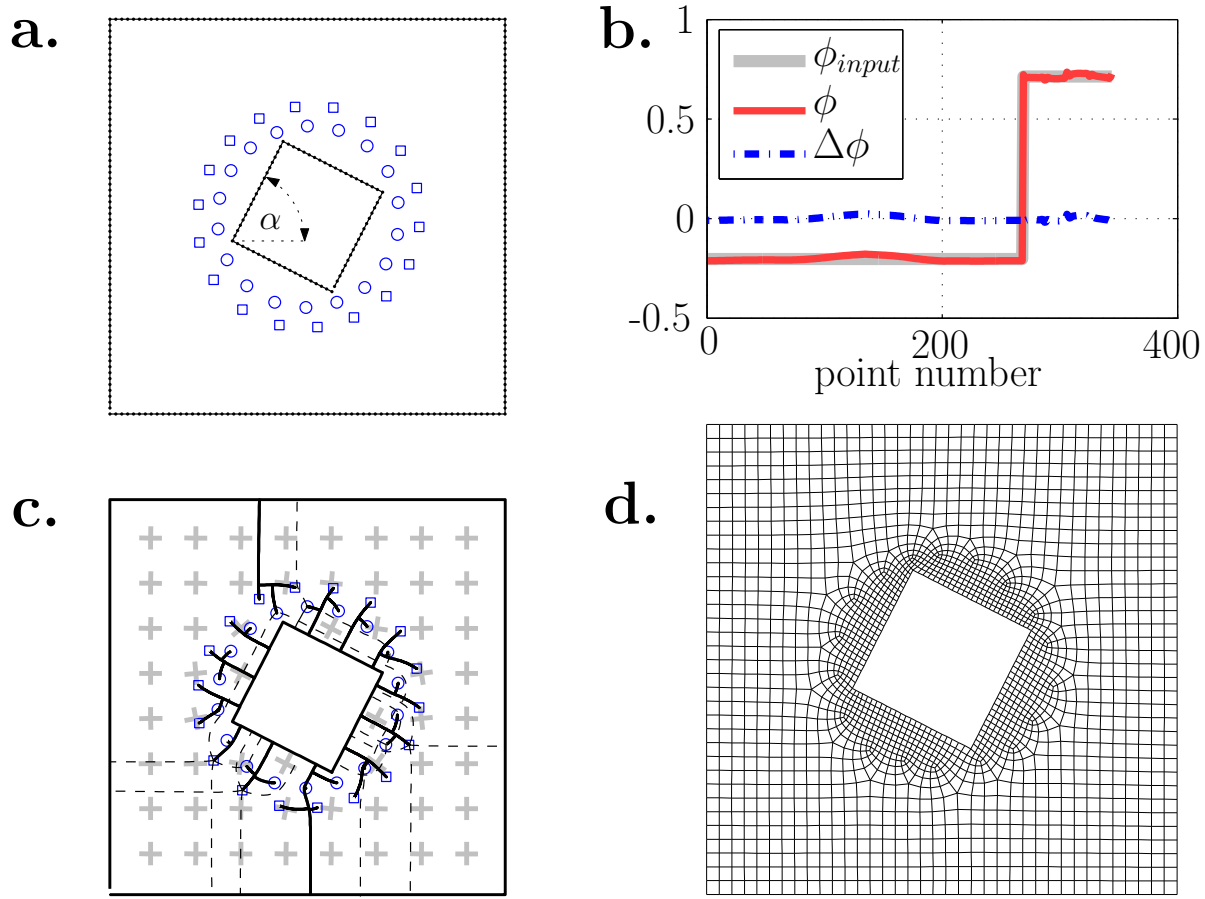}
\caption{(Color online) Mesh generation procedure: \textbf{a.} Domain boundary
(solid line) sampled at points (dots) as input to the IP algorithm. Resulting
charge distribution is plotted: $k=+1,-1$ cone-points are plotted with squares
and circles, respectively. \textbf{b.} $\phi$ obtained at boundaries. Points
1-268 correspond to the outer boundary component, the rest to the inner
component. Input $\phi$, obtained $\phi$ and their difference are plotted.
\textbf{c.} Crosses at selected points (crosses), cut-tree composed af the
boundary and star-geodesics (solid line), and selected additional
star-geodesics (dashed lines). \textbf{d.} The final mesh.}
\label{multi_diamond_var_res}
\end{center}
\end{figure}

\begin{figure}
[ptbh]
\begin{center}
\includegraphics[
height=5.4089in,
width=6.249in
]
{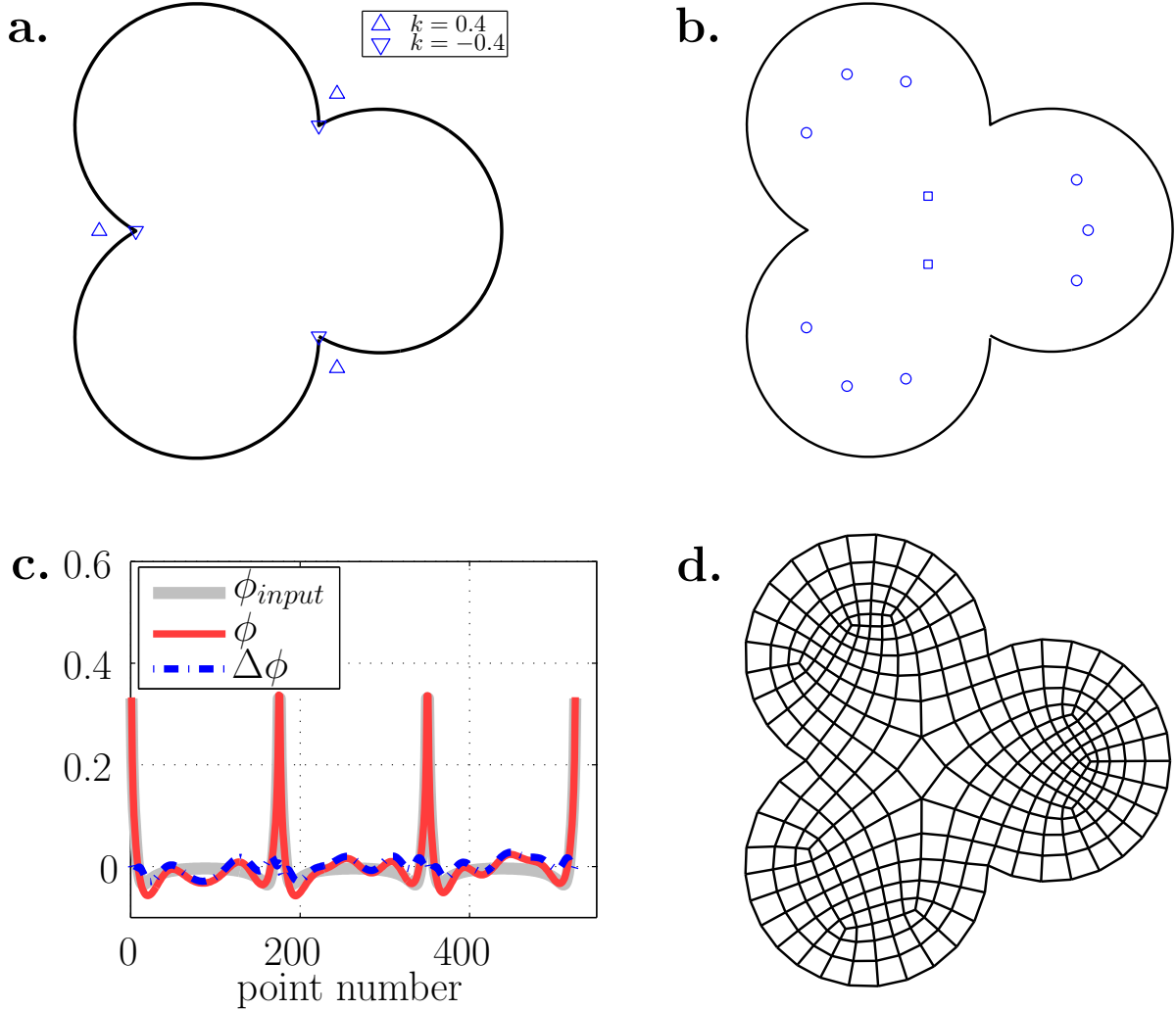}
\caption{(Color online) \textbf{a.} Boundary with adjustment of resolution
near junction points. \textbf{b.} Obtained source locations (blue).
\textbf{c.} Input $\phi$, obtained $\phi$ and their difference. \textbf{d.}
Final mesh.}
\label{fig:multi_ex2}
\end{center}
\end{figure}

\begin{figure}
[ptbh]
\begin{center}
\includegraphics[
height=5.3904in,
width=6.2733in
]
{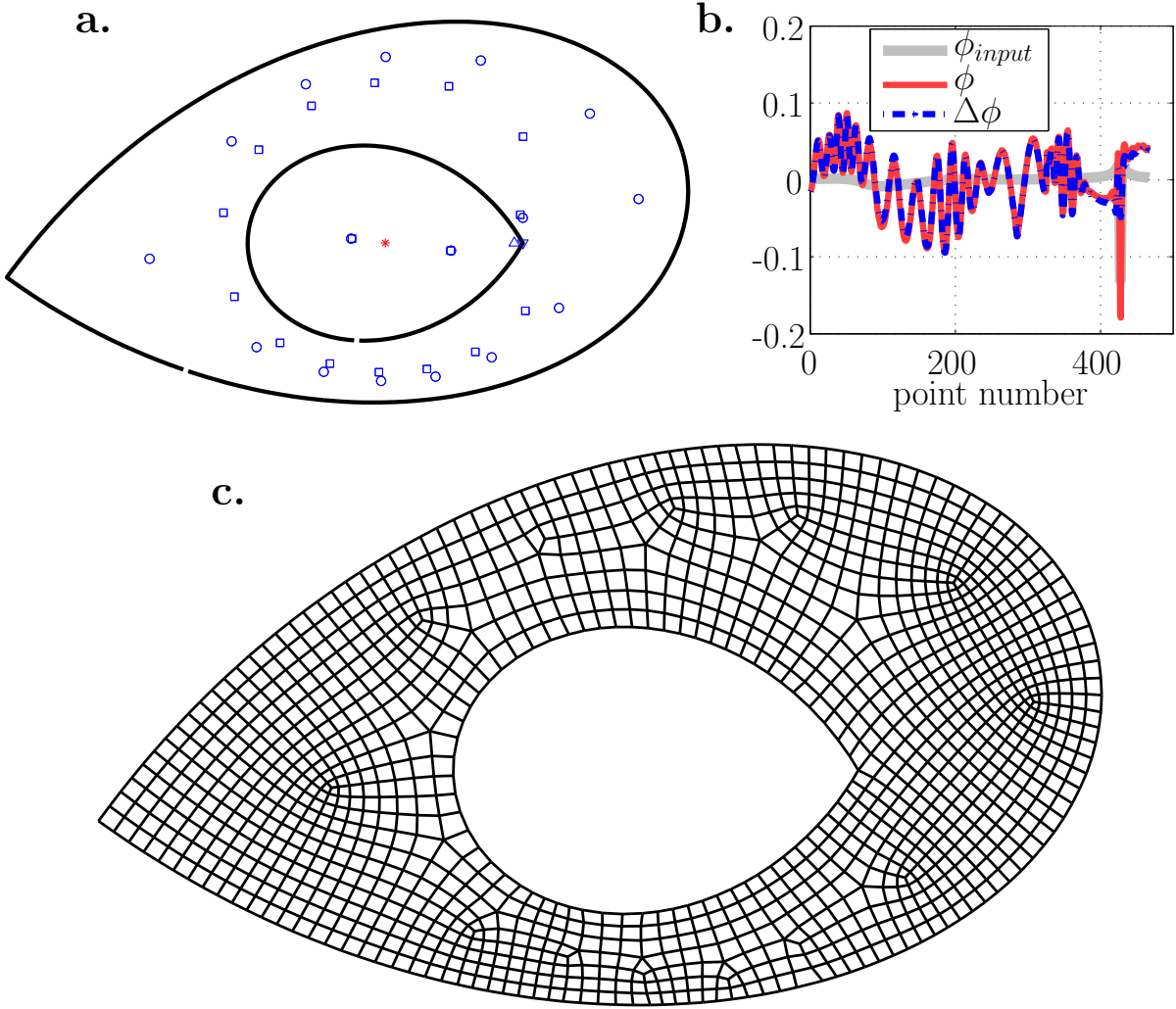}
\caption{(Color online) \textbf{a.} The boundary (solid line), sources to
adjust input resolution near junction (triangles), source distribution
(squares and circles), and inside source location (red star), used in removing
the harmonic part. \textbf{b.} Input $\phi$, obtained $\phi$ and their
difference. \textbf{c.} Final mesh.}
\label{fig:multi_ex3}
\end{center}
\end{figure}

\begin{figure}
[ptbh]
\begin{center}
\includegraphics[
trim=0.000000in 0.000000in 0.001224in 0.000000in,
height=5.3107in,
width=6.249in
]
{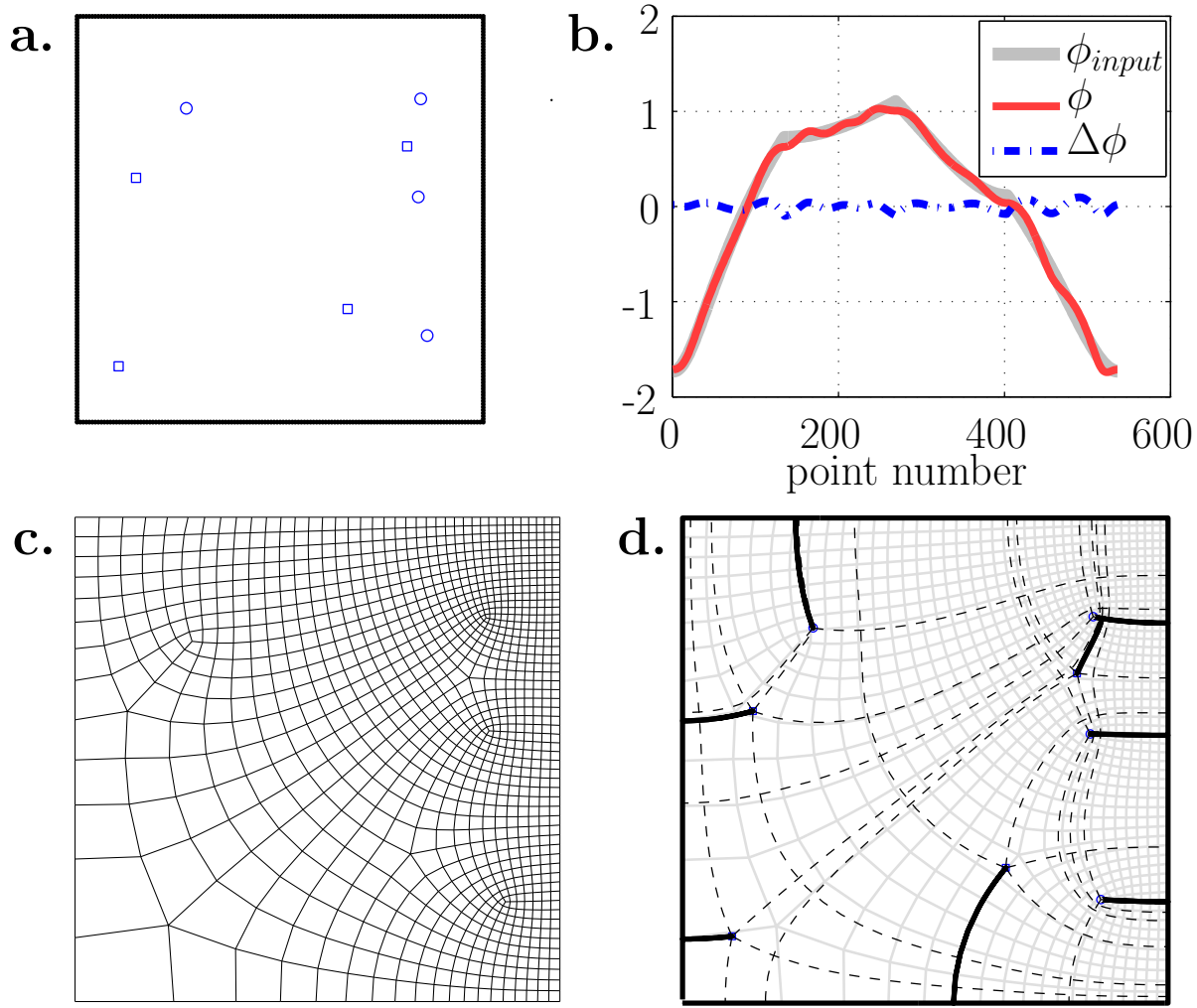}
\caption{(Color online) \textbf{a.} Boundary, source distribution (squares and
circles). \textbf{b.} Input $\phi$, obtained $\phi$ and their difference.
\textbf{c.} Final mesh. \textbf{d.} Final mesh (gray lines), overlaid with the
cut-tree (solid line) and additional star-geodesics (dashed lines).}
\label{fig:multi_ex4}
\end{center}
\end{figure}

In the first example, a domain with two boundary components is meshed, see
fig. \ref{multi_diamond_var_res},a. The side lengths of the outer and inner
squares are $8$ and $2.26$ respectively, centered at the origin, and the inner
square is rotated at angle $\alpha=0.35\pi$ radians. The ratio of input
resolution between outer and inner boundaries is $2.5$. The inner and outer
boundaries were sampled with $268$ and $76$ points, respectively, which were
used both for calculating the Cauchy integral, and for the sources' locations
calculation. Singularities of $\phi$ are not required at the junctions points,
since all inner angles are multiples of $\pi/2$ radians. As there is no flux
of $\nabla\phi$ through either boundary (because the boundaries' curvature is
$\kappa=0$, see eq. (\ref{eq:cond1}), and $\phi$ is regular at the junctions),
the total charge inside is zero (see remark \ref{remark:divergence}). Hence an
equal number of $k=+1$ and $k=-1$ charges were used. Fig.
\ref{multi_diamond_var_res},a., shows the domain boundary, together with the
calculated sources' locations. Squares mark sources with $k=+1$, and circles
sources with $k=-1$. The distribution is composed of $18$ sources of each
type, i.e. a total of $36$ sources.\ The value of $\phi$ obtained by solving
the Poisson equation with this\ charge distribution is shown in fig.
\ref{multi_diamond_var_res},b., and compared to the input $\phi$ requested,
denoted by $\phi_{input}$. The difference $\Delta\phi\equiv\phi-\phi_{input}$
is also plotted. For visual clarity, the average of the $\phi_{input}$ was
subtracted from both $\phi$ and $\phi_{input}$ (this is just a scaling of the
resolution by a constant factor); this was also done in the resolution
comparisons below, in fig. (\ref{fig:multi_ex2}),c., fig. (\ref{fig:multi_ex3}
),b. and fig. (\ref{fig:multi_ex4}),c.. The number of sources was chosen to be
the smallest for which $\Delta\phi<0.1$ at all boundary points. This criterion
is used for choosing the number of sources is also used in the following
examples. The cut-tree used in creating the final mesh is shown in Fig.
\ref{multi_diamond_var_res},c., and the final mesh in shown Fig.
\ref{multi_diamond_var_res},d.

The domain in the second example is the union of three unit radius circles
which pass through the origin, see fig. \ref{fig:multi_ex2},a. The boundary
has three junctions, all with the same inner angle, $\theta_{in}=5\pi/3$.
Boundary alignment requires $\phi$ to be singular at the junctions,
corresponding to sources with charge $Q=-\pi/5$, or $k$-value $k=Q/(\pi
/2)=-2/5$, see eq. (\ref{eq:junc_charge}),(\ref{eq:junc_incidence}). As
explained in section \ref{sec:input_near_junc}, singular functions can be
added the input resolution in order to avoid clustering of sources near the
junction points. In this example, we compose the input $\phi$ from a sum of
pairs of source terms, with $k$-values as marked in fig. \ref{fig:multi_ex2}
,a.. The pairs of nearby sources with opposite charge create the desired
singularities at the junctions, and have only a small effect at distances
large compared with the distance between the sources of each pair.

We can now calculate the total flux of $\phi$ through the boundary in this
example. The (singular) flux through each junction is $\pi/6$ (positive value
when $\nabla\phi$ is directed outward). The flux through each of the three
arcs is $\int\partial_{n}\phi ds=-\int\partial_{s}\theta ds=-\Delta\theta$
along the arc, and since for each arc $\Delta\theta=4\pi/3$, the total flux
through the boundary is $3\cdot\left(  \pi/6\right)  +3\cdot\left(
-4\pi/3\right)  =-7\left(  \pi/2\right)  $. This is in agreement with the
formula given in remark \ref{remark:flux_calc}. Therefore, the sum of
$k$-values of the sources inside the domain should be $-7$, so $n_{-}=n_{+}
+7$. Hence, $k_{in}$ of eq. (\ref{eq:source_in_D}) was $-7$; $z_{in}=1$ was
used. The lowest number of sources for which $\Delta\phi<0.1$, shown in fig.
\ref{fig:multi_ex2},b., has $n_{+}=2$, $n_{-}=9$. The boundary was sampled at
$525$ points.

The boundary in the third example is composed of two boundary components, each
with a single junction point. The inner angles at the inner and outer boundary
junctions are $\theta_{in}=4\pi/3,\pi/2$ respectively, so a singularity of
$\phi$ will form at the inner boundary junction. As in the previous example,
we add a singular function to the input $\phi$, by adding a pair of
source\ terms, one at the junction point with $k=1/2$, as obtained from eq.
(\ref{eq:junc_charge}),(\ref{eq:junc_incidence}) with $\theta_{in}=4\pi/3$. No
inner source term is required, and from the flux trough the hole boundary it
follows that $k_{hole}=-3$. After subtracting the junction source term with
$k=1/2$ \ (see section \ref{sec:junc_points}) a source with total charge of
$k_{hole}=-3-1/2$ is added inside the hole, see fig. \ref{fig:multi_ex3},a..
The inner and outer boundary components are sampled at $325$ and $142$ points,
respectively. The distribution with $n_{+}=n_{-}=16$, the smallest number of
sources for which $\Delta\phi<0.1$, is plotted at fig. \ref{fig:multi_ex3},a.,
with squares and circles representing charges as in previous examples. Two
pairs of $k=\pm1$ charges are located inside the inner boundary, i.e., outside
the domain. Their constitute a correction to the homogeneous $\phi_{L}$, which
is very small, since in each pair the opposite charges almost overlap. The
obtained $\phi$ reconstruction is given in fig. \ref{fig:multi_ex3},b.. The
final mesh, fig. \ref{fig:multi_ex3},c., contains $26$ irregular vertices,
$13$ of each type. This is because two pairs of sources lay outside the
domain, and another pair (the pair of sources closest to the inner junction)
was \textquotedblleft eliminated\textquotedblright\ in the creation of the
final mesh (section \ref{sec:create_final}): the two opposite charges where
united, giving a zero total charge.

In the final example the boundary of the square $[-2,2]\times\lbrack-2,2]$ was
assigned a varying\ input resolution proportional to:
\[
resolution\propto1.5+(x+2)^{2}+\frac{(y+2)^{2}}{2}\text{.}
\]

The resolution thus varies by a factor of $17$ within the domain. As always,
eq. (\ref{eq:phi_from_res}), $\phi=\ln\left(  resolution\right)  $. The total
charge is zero, so $n_{+}=n_{-}$. The boundary was sampled at $536$ points. A
source distribution with $\Delta\phi<0.1$ was obtained with just $8$ charges,
i.e. with $n_{+}=n_{-}=4$. Fig. \ref{fig:multi_ex4},a. shows the source
distribution, Fig. \ref{fig:multi_ex4},b. shows the obtained $\phi$ vs. the
input $\phi$, and the final mesh is shown in Fig. \ref{fig:multi_ex4},c.. It
is instructive to compare the star-geodesics and cut-tree with the final mesh.
The two are overlaid in Fig. \ref{fig:multi_ex4},d. (this would create a
\textquotedblleft crowded\textquotedblright\ appearence when more sources are
present). The black solid line traces the cut-tree, the dashed lines are
additional star-geodesics, and the mesh is drawn with the background in gray lines.

\bigskip

Cell shape quality was measured using a variant of the quality measure $\beta$
\cite{Lo}, as defined in \cite{qmorph}. $\beta=1$ represents a square cell,
while $\beta=0$ represents a cell with an inner angle of $\pi$. Cells with a
high aspect ratio are also given low $\beta$-values. Table 1 shows the minimum
and average $\beta$ values, and the total number of cells, for the examples in
Fig. \ref{multi_diamond_var_res}-\ref{fig:multi_ex4}. For each example, the
tabulated information is given for the mesh shown in the corresponding figure,
and for a finer mesh of the same domain, where the resolution function was
doubled (prodcing a mesh with about four times the number of cells). Note that
multiplying the resolution by a factor amounts to adding a constant to $\phi$,
which does not change the source distribution obtained from the IP algorithm,
as this constant, which is part of $h\left(  z\right)  $, is readily removed
when $h\left(  z\right)  $ is subtracted.

\begin{center}
\bigskip$
\begin{tabular}
[c]{|c|c|c|c|}\hline
& num. cells & min. $\beta$ & avg. $\beta$\\\hline
Fig. \ref{multi_diamond_var_res} & 1898 & 0.576 & 0.945\\\hline
& 7630 & 0.627 & 0.975\\\hline
Fig. \ref{fig:multi_ex2} & 316 & 0.603 & 0.901\\\hline
& 1036 & 0.620 & 0.942\\\hline
Fig. \ref{fig:multi_ex3} & 1233 & 0.506 & 0.933\\\hline
& 4895 & 0.562 & 0.958\\\hline
Fig. \ref{fig:multi_ex4} & 1062 & 0.626 & 0.964\\\hline
& 6059 & 0.684 & 0.987\\\hline
\end{tabular}
$

$\ \ $

Table 1: Mesh statistics.
\end{center}

\section{Conclusions and Future work\label{sec:conclusions}}

An unstructured quadrilateral mesh generation scheme in planar domains was
presented. The method rests on a theoretical foundation, linking the mesh
generation problem with the Inverse Poisson (IP) problem. An IP solution
algorithm is presented, whose output is interpreted as the location and type
(degree) of irregular vertices in the domain. The continuum fields obtained,
describing mesh resolution and directionality, are conformal everywhere except
on the irregular vertices, and fit the required input properties at the
boundary, or at other user-defined locations. An algorithm for creating a
valid final mesh is also presented. Example meshes feature irregular vertices
where they are needed, in combination with highly regular regions where possible.

Directions for future work include\ more sophisticated methods for solving the
rational function interpolation equations, and for constructing the final
mesh. The relations between conformal unstructured mesh generation, the IP
problem and rational function interpolation raise many research questions.
These may lead to a deeper understanding of the properties of high quality
meshes, and to better algorithms for creating them.

\bigskip

\textbf{Acknowledgements.} The author would like to thank Mirela Ben-Chen,
Shlomi Hillel, Dov Levine, Yair Shokef and Vincenzo Vitelli for helpful
discussions and critical reading of the manuscript.

\section{Appendix A: Star-geodesic directions}

A cross is defined at every point $p$ of the domain, that is not a singularity
of the function $\phi$. Geodesic curves that start at $p$ can be drawn in all
four directions of the cross. By the definition of a cross-field, such a
geodesic will be aligned with the cross-field everywhere along the curve. If
$p$ is a singular point, with $k$-value~$k_{p}\neq0$, a cross is not defined
at $p$, but there are $4+k_{p}$ geodesics that are incident on $p$ and follow
the cross-field directions elsewhere on the curve. These will be called
\emph{star-geodesics}. For example, the geodesics drawn in fig.
\ref{multi_diamond_var_res},c.\ and \ref{fig:multi_ex4},d.\ are star-geodesics.

\begin{figure}
[ptb]
\begin{center}
\includegraphics[
height=2.093in,
width=2.7928in
]
{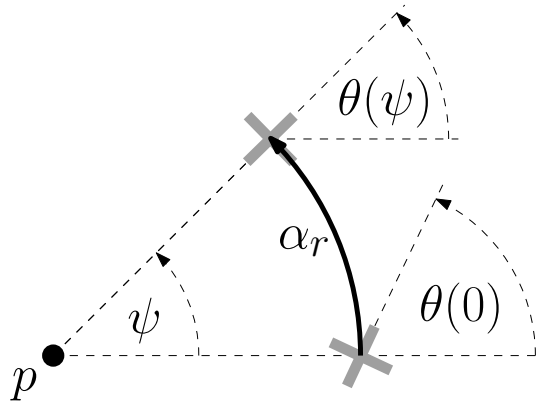}
\caption{Calculating the star-geodesics' directions.}
\label{fg:star_directions}
\end{center}
\end{figure}

Denote the angle from the $x$-axis around $p$ by $\psi$, and the cross
direction when $p$ is approached from direction $\psi$ by $\theta\left(
\psi\right)  $, see fig. \ref{fg:star_directions}. To calculate the directions
in which star-geodesics emanate from $p$, we first calculate the cross when
$p$\ is approached from some direction, e.g. the positive $x$-axis. This can
be done using eq. (\ref{eq:theta_change_par_tranport}) along a curve from the
boundary to $p$ which approaches $p$ from the direction $\psi=0$. Once
$\theta\left(  0\right)  $ is known, $\theta\left(  \psi\right)  $ for any
$\psi$ can be calculated by using eq. (\ref{eq:theta_change_par_tranport})
along a small circular arc around $p$ at radius $r$, $\alpha_{r}$. The
singularity term at $\phi$ is $\frac{k}{4}\ln\left\vert r\right\vert $, and
according to eq. (\ref{eq:theta_change_par_tranport})
\begin{equation}
\theta\left(  \psi\right)  =\theta\left(  0\right)  +\int_{\alpha_{r}}
\frac{\partial\phi}{\partial n}ds=\theta\left(  0\right)  +\frac{\partial\phi
}{\partial r}\psi r=\theta\left(  0\right)  +\frac{k}{4r}\psi r=\theta\left(
0\right)  +\frac{k}{4}\psi.\label{eq:app_a_th_calc}
\end{equation}
Note that contributions to $\phi$ that are regular at $p$ do not affect
$\theta\left(  \psi\right)  $ for $r\rightarrow0$.

The star-geodesic directions are those for which the cross is directed along
the ray from $p$, or
\begin{equation}
\theta\left(  \psi\right)  =\psi+n\frac{\pi}{2},\label{eq:app_a_star_cond}
\end{equation}
with $n\in
\mathbb{Z}
$. Substituting eq. (\ref{eq:app_a_th_calc}) into eq (\ref{eq:app_a_star_cond}
) and rearranging, we find
\[
\psi=\frac{\theta\left(  0\right)  -n\frac{\pi}{2}}{1-\frac{k}{4}}\text{.}
\]
It is easy to show that there are exactly $4+k$ different $\psi$-values for
which this equation is fulfilled. They are equally distributed around $p$.

\end{document}